\documentclass[prb,aps,twocolumn]{revtex4}

\usepackage{graphicx}
\usepackage{dcolumn}
\usepackage{bm}
\usepackage{amssymb}

\begin{document}


\title{Pyrochlore antiferromagnet with antisymmetric exchange
interactions: \\ critical behavior and order from disorder}

\author{Gia-Wei Chern}
\affiliation{Department of Physics, University of Wisconsin,
Madison, Wisconsin 53706, USA}

\date{\today}

\begin{abstract}
We investigate the nature of phase transitions induced by
Dzyaloshinskii-Moriya (DM) interactions on a classical pyrochlore
antiferromagnet. For both symmetry-allowed antisymmetric exchange
interactions, the macroscopic degeneracy due to geometrical
frustration is relieved and a long-range magnetic order appears in
the ground state. We find an Ising-like phase transition in the case
of direct DM interaction which selects a doubly degenerate all-in
all-out magnetic order. In the presence of indirect DM coupling, the
magnet undergoes an XY-like phase transition into a state with
broken $Z_6$ symmetry. We show that the critical behavior in both
cases is modified due to the constrained spin fluctuations in the
correlated Coulomb phase. We also demonstrate an interesting order
from disorder phenomenon where the system switches between two
distinct types of broken $Z_6$ symmetry.
\end{abstract}

\maketitle

\section{Introduction}

\label{sec:intro}

Geometrical frustration refers to the inability of spins to satisfy
conflicting interactions simultaneously due to the lattice
connectivity. \cite{moessner06} Magnets with strong geometrical
frustration exhibit a variety of unusual ground states, elementary
excitations, and phase transitions. One of the most intensively
studied frustrated systems is the Heisenberg antiferromagnet on the
{\em pyrochlore} lattice. Monte~Carlo simulations with interactions
restricted to nearest-neighbor spins showed that the magnet remains
disordered even at temperatures well below the energy scale of
exchange constant. \cite{moessner98} Contrary to the uncorrelated
paramagnetic state, motions of spins in this so~called Coulomb phase
are subject to a set of local constraints. In the continuum
approximation, these local constraints translate to the Gauss's law
for a fictitious magnetic field. \cite{hermele04,isakov04,henley05}

Many interesting properties of the Coulomb phase can be traced to
the macroscopic degeneracy in the classical ground state of the
pyrochlore magnet. Indeed, minimization of the nearest-neighbor
exchange interaction requires that the total spin on every
tetrahedron be zero, which still leaves an extensive number of
unconstrained degrees of freedom. \cite{moessner98} This degeneracy
in turn makes the magnet susceptible to small perturbations such as
anisotropies, dipolar interaction,~etc. In general, the frustration
is relived and some sort of long-range order is selected by the
dominant perturbations.

Phase transitions taking place in the highly constrained Coulomb
phase also exhibit interesting, and sometimes peculiar, behavior. A
recurring theme in experiments and numerical simulations has been
the appearance of first-order phase transitions, despite that a
Landau-type analysis of the symmetry-breaking phase would otherwise
predict a continuous one. Discontinuous phase transitions have been
observed in pyrochlore antiferromagnet perturbed by magnetoelastic
coupling,\cite{lee00,chung05} long-range dipolar interaction,
\cite{melko01,cepas05} exchange anisotropy,
\cite{pickles08,saunders08} and further-neighbor exchange
interactions. \cite{tsuneishi07,chern08} It is known that strong
fluctuations of the order parameter could induce a first-order
transition, a scenario frequently occurs when the number of
components of the order parameter is greater than four.
\cite{brezin74,bak76} This is indeed the case for most of the phase
transitions mentioned above.

The dipolar spin correlations stemming  from the strong local
constraints also affect the nature of phase transitions from the
Coulomb phase. \cite{hermele04,isakov04,henley05} As is well known,
the long-range dipolar interaction can change the universality class
of a magnetic phase transition, modifying its critical behavior to
that of a model with short-range interactions in higher spatial
dimensions. \cite{larkin69,aharony73} In a recent study on the
effects of the exchange anisotropy, it was demonstrated that N\'eel
ordering from the Coulomb phase indeed displays a critical behavior
similar to that of a uniaxial dipolar antiferromagnet.
\cite{pickles08}

As most broken symmetries observed in pyrochlore lattice are rather
complex, it is desirable to study phase transitions which can be
described by simple order parameters. In this paper, we investigate
critical behavior of magnetic ordering due to asymmetric exchange
coupling, or the Dzyaloshinskii-Moriya (DM) interaction.
\cite{dz,moriya} The relevant order parameters have either a $Z_2$
or $Z_6$ symmetry, depending on the sign of the DM interaction.
Specifically, we study the classical spin Hamiltonian:
\begin{eqnarray}
    \label{eq:Hdm}
    \mathcal{H} = J \sum_{\langle ij \rangle} \mathbf S_i\cdot\mathbf S_j +
    \sum_{\langle ij \rangle} \mathbf D_{ij}\cdot\left(\mathbf
    S_i\times \mathbf S_j\right),
\end{eqnarray}
where the two terms represent the isotropic exchange and the DM
interactions, respectively, between nearest-neighbor spins. In
addition to being an interesting spin model in itself, recent
studies have revealed the important role of DM interaction in the
formation of magnetic spirals \cite{chern06,ot} and the magnon Hall
effect \cite{onose10} observed in the pyrochlore magnets.

By applying the so~called Moriya rules, \cite{moriya} the possible
forms of the coupling vectors $\mathbf D_{ij}$ were determined by
Elhajal {\em et al.} up to a multiplicative constant.
\cite{elhajal05} Depending on the sign of this constant, the
corresponding DM terms are called direct and indirect DM
interactions. The ground states for both forms of the DM interaction
were also obtained in Ref.~\onlinecite{elhajal05}.  In particular,
it was pointed out that indirect DM interaction leaves a
continuously degenerate ground state at the mean-field level.
Monte~Carlo simulations, however, showed that a six-fold degenerate
ground state with coplanar spins was selected by thermal
fluctuations. \cite{elhajal05,canals08} Although their preliminary
simulations indicated a continuous magnetic ordering for both types
of DM interaction, the authors did not touch upon the nature of the
phase transitions.

\begin{figure}
\centering
\includegraphics[width=0.9\columnwidth]{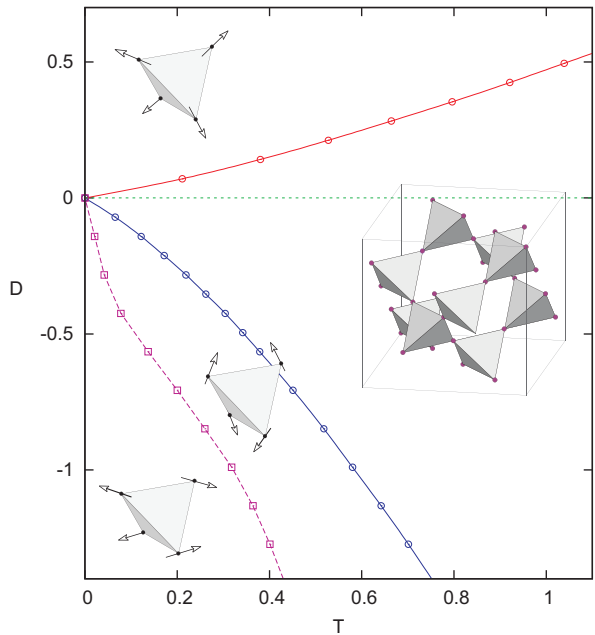}
\caption{\label{fig:phase} Phase diagram of Heisenberg
antiferromagnet with Dzayloshinskii-Moriya interaction on pyrochlore
lattice (shown in the inset). The ground state of direct DM
interaction $(D>0)$ consists of all-in all-out tetrahedra. The
ordered phase of indirect DM interaction $(D<0)$ contains two
distinct ordered states with broken $Z_6$ symmetry: noncoplanar and
orthogonal structures at high and low $T$, respectively. The solid
and dashed lines denote second-order transition and crossover lines,
respectively.}
\end{figure}

In this paper we performed detailed finite-size scaling to study the
critical behavior of model (\ref{eq:Hdm}). The phase diagram
obtained from our extensive Monte Carlo simulations is shown in
Fig.~\ref{fig:phase}. Consistent with previous results,
\cite{elhajal05} we found that the direct DM term is minimized by an
all-in all-out spin structure, whereas the indirect DM interaction
selects orthogonal magnetic orders as $T \to 0$. By carefully
characterizing the symmetry-breaking phase in the indirect DM model,
we find a {\em new} intermediate regime characterized by a magnetic
order with noncoplanar spins. A crossover line separates the two
different types of broken $Z_6$ symmetry.

Due to the $Z_2$ symmetry of the doubly degenerate all-in all-out
spin structure, magnetic ordering induced by direct DM interaction
($D>0$) is expected to be Ising-like. The critical exponents
obtained from finite-size scaling indicate that the transition is
indeed close to 3D Ising universality class. We show that the
constrained spin fluctuations in the Coulomb phase result in
noticeable deviation especially in exponents $\beta$ and $\gamma$.

The case of indirect DM interaction ($D<0$) is more complicated.
Upon lowering the temperature, the magnet undergoes an XY-like
transition into a phase with a broken $Z_6$ symmetry. We introduce a
doublet order parameter $\bm\zeta = (\zeta_x, \zeta_y)$ such that
the six coplanar ground states correspond to the six corners
$\theta_{\zeta} = n\pi/3$ of a hexagon in the $\zeta$-domain; here
$n$ is an integer. Interestingly, below the critical temperature the
magnet enters an ordered state characterized by $\theta_{\zeta} =
(n+1/2)\,\pi/3$; the corresponding magnetic order consists of
noncoplanar spins. As temperature is further lowered the system
gradually evolves toward the coplanar magnetic structure. The
crossover between the two ordering patterns does not break further
symmetries.

The continuous degeneracy of the noncoplanar ground states observed
in Ref.~\onlinecite{elhajal05} manifests itself as an O(2) symmetry
of the doublet  order parameter $\bm\zeta$. Consequently, both types
of broken $Z_6$ symmetry are entirely due to the entropic effect, or
so~called order-from-disorder. By explicitly computing the magnon
contribution to the free energy, we demonstrate that the above
crossover phenomenon is indeed dictated by thermal fluctuations.

The remainder of the paper is organized as follows. In
Sec.~\ref{sec:dm}, we introduce three antiferromagnetic order
parameters to characterize the spin configurations in the ordered
phase. The critical behavior of direct and indirect DM interactions
is discussed in Secs.~\ref{sec:direct-dm} and \ref{sec:indirect-dm},
respectively. A quantitative analysis of the crossover between the
two distinct types of broken $Z_6$ symmetry is presented in
Sec.~\ref{sec:crossover}. Finally we conclude with a discussion of
these results in Sec.~\ref{sec:conclusion}.

\section{Ground-state magnetic orders}
\label{sec:dm}

From the symmetry viewpoint, the DM interaction is allowed on the
pyrochlore lattice since the nearest-neighbor bonds are not
cnetrosymmetric. \cite{moriya} In Moriya's original theory, the DM
term stems from the relativistic spin-orbit interaction.
\cite{moriya} Regardless of its microscopic origin, the high
symmetry of the pyrochlore lattice places strict constraints on the
direction of the coupling vectors. \cite{elhajal05} For a bond
oriented along the $[110]$ direction, the vector $\mathbf D_{ij}$
points along $[1\bar{1}0]$: $\mathbf D_{ij} = (\pm D, \mp D,
0)/\sqrt{2}$. The value of $\mathbf D_{ij}$ on any other bond can
then be found through symmetry transformations.

\begin{figure}
\centering
\includegraphics[width=0.95\columnwidth]{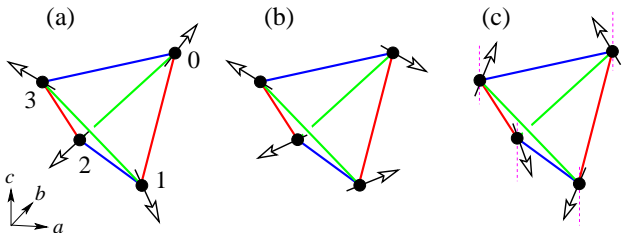}
\caption{\label{fig:tetra} Magnetic states of a tetrahedron with
zero total spin $\mathbf M = 0$: (a) all-out, (b) orthogonal, and
(c) noncoplanar structures. The three antiferromagnetic vectors
$\mathbf L_1$, $\mathbf L_2$, and $\mathbf L_3$ measure the
staggered magnetizations between the two red, green, and blue bonds,
respectively.}
\end{figure}

The expression of the DM interaction could be further simplified
using the antiferromagnetic order parameters $\mathbf L_1 = (\mathbf
S_0 + \mathbf S_1 - \mathbf S_2 - \mathbf S_3)/4S$,\, \cite{chern06}
and so on (see Fig.~\ref{fig:tetra} for labeling of the spins),
which measure the staggered magnetizations of a tetrahedron. For
completeness we also define the ferromagnetic order parameter:
$\mathbf M = \sum_{i=0}^3\mathbf S_i/4S$. In terms of these
variables, Hamiltonian~(\ref{eq:Hdm}) can be recast into a sum over
tetrahedra $\mathcal{H} = \sum_{\boxtimes} \mathcal{H}_{\boxtimes}$,
where the Hamiltonian of a single tetrahedron is
\begin{eqnarray}
  \label{eq:H2}
  \mathcal{H}_{\boxtimes} &=& 8JS^2 \left|\mathbf M\right|^2
  -4\sqrt{2}DS^2 \bigl(\hat{\bf a}\cdot\,{\bf L}_2\times{\bf L}_3
  \nonumber \\
  & & \quad +\,\hat{\bf b}\cdot\,{\bf L}_3\times{\bf L}_1
  +\hat{\bf c}\cdot\,{\bf L}_1\times{\bf L}_2\bigr).
\end{eqnarray}
Minimization of the exchange energy requires $\mathbf M = 0$ on
every tetrahedron, which still leaves a macroscopic degeneracy. The
remaining degrees of freedom encoded in the three staggered vectors
are determined by minimization of the DM term.

Since the vanishing of $\mathbf M$ makes the three staggered
magnetizations orthogonal to each other and imposes a constraint on
their lengths: $\sum_{i=1}^3 \left| \mathbf L_i \right|^2 = 1$,
\cite{chern06} we parameterize the antiferromagnetic order
parameters as $\mathbf L_i = \phi_i\hat\mathbf n_i$, where the three
$\hat\mathbf n_i$ form an orthogonal triad and $\phi_i > 0$ denote
their lengths. Defining a `handedness' for the triad: $\chi \equiv
\hat\mathbf n_1\cdot\hat\mathbf n_2\times \hat\mathbf n_3 = \pm 1$,
the DM term becomes
\begin{eqnarray}
    \label{eq:dm2}
    -\chi DS^2 \bigl(\hat\mathbf n_1\cdot\hat\mathbf
    a\,\phi_2\phi_3 + \hat\mathbf n_2\cdot\hat\mathbf b\,\phi_3\phi_1
    + \hat\mathbf n_3\cdot\hat\mathbf c\,\phi_1\phi_2\bigr).
\end{eqnarray}
Note that inversion of the triad $\hat\mathbf n_i \to -\hat\mathbf
n_i$ also changes the chirality $\chi \to -\chi$, reflecting the
time-reversal symmetry of the DM interaction. In the following, we
focus on the case of a right-handed triad with $\chi = +1$.

For $D>0$ the DM interaction is minimized first by aligning the
triad to the principal crystal axes such that $\hat\mathbf n_1 =
\hat\mathbf a$, and so on. Minimization of the resultant expression
subjecting to the constraint $\sum_{i=1}^3\phi_i^2 = 1$ yields a
ground state with $\phi_1=\phi_2 =\phi_3 = 1/\sqrt{3}$. The
staggered magnetizations are given by
\begin{eqnarray}
    \label{eq:all-out}
    \mathbf L_1 = (\chi/\sqrt{3})\,\hat\mathbf a,\,\,\,\,
    \mathbf L_2 = (\chi/\sqrt{3})\,\hat\mathbf b,\,\,\,\,
    \mathbf L_3 = (\chi/\sqrt{3})\,\hat\mathbf c,
\end{eqnarray}
with $\chi = \pm 1$ corresponding to the all-out
[Fig.~\ref{fig:tetra}(a)] and all-in structures, respectively.

The magnet remains frustrated in the presence of the indirect DM
interaction, despite a significantly reduced degeneracy. The
frustration comes from the fact that absolute minimum of the DM
energy with $D<0$ and $\chi = +1$ requires the unit vectors
$\hat\mathbf n_i$ be antiparallel to the corresponding principal
directions, e.g. $\hat\mathbf n_1 = -\hat\mathbf a$. However, such a
complete anti-alignment is impossible for two right-handed triads.
Numerical minimization yields two inequivalent classes of
continuously degenerate ground states. The first set consists of
orthogonal spins lying in the plane perpendicular to one of the
three cubic axes. The energy of a coplanar state is invariant with
respect to uniform rotations of spins about the corresponding cubic
axes. For spins perpendicular to the $c$ axis, the magnetic state of
the tetrahedron is described by
\begin{eqnarray}
    \label{eq:orthogonal}
    & &\mathbf L_1 = \bigl(\cos\phi\,\hat\mathbf a +
    \sin\phi\,\hat\mathbf b\bigr)/\sqrt{2}, \nonumber \\
    & & \mathbf L_2 = \bigl(\sin\phi\,\hat\mathbf a -
    \cos\phi\,\hat\mathbf b\bigr)/\sqrt{2}, \nonumber \\
    & & \mathbf L_3 = 0.
\end{eqnarray}
where the angle $\phi$ describes the uniform rotation about $c$
axis. As $T \to 0$, thermal fluctuations select coplanar states with
$\phi = 0$ or $\pi$. \cite{canals08} The orthogonal structure with
spins pointing along the $[110]$ and $[1\bar 10]$ directions ($\phi
= 0$) is shown in Fig.~\ref{fig:tetra}(b).

The second set of ground states consists of noncoplanar spins
described by order parameters
\begin{eqnarray}
    \label{eq:non-coplanar}
    \mathbf L_1 &=& \cos(\theta-\pi/4)\,\cos\Theta\,\hat\mathbf a, \nonumber
    \\
    \mathbf L_2 &=& \sin(\theta-\pi/4)\,\cos\Theta\,\hat\mathbf b, \nonumber
    \\
    \mathbf L_3 &=& -\sin\Theta\,\hat\mathbf c,
\end{eqnarray}
where $\Theta = \arctan\left(\sqrt{2}\sin\theta\right)$ and $\theta$
is a continuous parameter. A tetrahedron with $\theta = 0$
corresponds to orthogonal spins lying in the $ab$ plane. Starting
from $\theta = 0$, one can reach the other two inequivalent
orthogonal states at $\theta = \pm \pi/4$, where the spins are
perpendicular to $a$ and $b$ axes, respectively. As will be
discussed in Sec.~\ref{sec:crossover}, order by disorder at finite
temperatures prefers the noncoplanar spins with, e.g. $\theta =
\pi/2$, shown in Fig.~\ref{fig:tetra}(c).

As first noted in Ref.~\onlinecite{colon}, the coarsed-grained
antiferromagnetic order parameters are just the three components of
a flux or polarization field introduced in
Refs.~\onlinecite{isakov04} and \onlinecite{henley05} to describe
the correlated Coulomb phase. This correspondence thus implies that
the coarsed-graind staggered fields also obey a dipolar correlation
at large distances. A continuum theory of the antisymmetric
pyrochlore antiferromagnet is outlined in
Appendix~\ref{app:continuum}.

Since the DM term does not depend explicitly on $\mathbf M$, a
remarkable feature of Hamiltonian~(\ref{eq:H2}) is that its ground
state is independent of the antiferromagnetic exchange~$J$. However,
spin fluctuations in the $J=0$ and $J \to \infty$ (Coulomb phase)
limits are quite different. In the latter case, minimization of the
$J$ term requires spin fluctuations satisfy
$\sum_{i\in\boxtimes}\delta\mathbf S_i \approx 0$, which in turn
result in a dipolar spin correlation at large distances. In the
following, we compare the universality class of magnetic ordering in
the two limiting cases and show that the critical behavior is indeed
modified by the constrained fluctuations in the Coulomb phase.

\begin{figure}[b]
\includegraphics[width=1.\columnwidth]{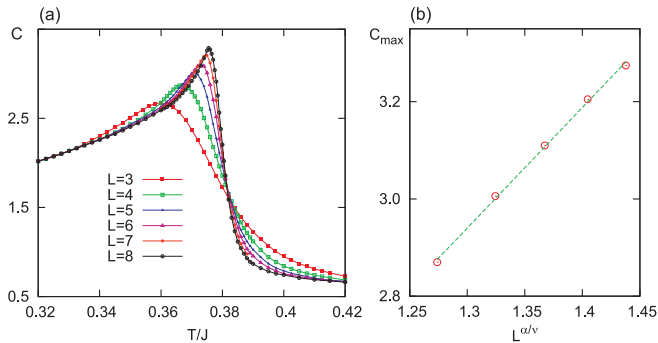}
\caption{\label{fig:d-dm1} Monte Carlo simulations for coupling
constants $D = 0.14J$. (a) Specific heat $C = \left(\langle
E^2\rangle - \langle E\rangle^2\right)/NT^2$ as a function of
temperature $T$ for various system sizes. (b) The maximum of the
specific heat $C_{\rm max}$ as a function of $L^{\alpha/\nu}$, where
$\alpha/\nu = 0.175$. The number of spins $N = 16 L^3$ for a system
with linear size $L$.}
\end{figure}

\section{Ising-like phase transitions: direct DM interaction}
\label{sec:direct-dm}

The direct DM interaction removes the degeneracy completely and
selects a doubly degenerate all-in all-out ground state. The magnet
undergoes a continuous phase transition at a temperature $T_c \sim
\mathcal{O}(D)$ as demonstrated by the specific heat curves shown in
Fig.~\ref{fig:d-dm1}(a). The peak of the specific heat becomes
sharper with increasing system size. However, finite-size scaling of
the specific heat is rather difficult due to a nonzero regular
component. Instead, by plotting the maximum of heat capacity vs the
linear size of the system, we found a power-law dependence $C_{\rm
max} = C_0 + {\rm const}\times L^{\alpha/\nu}$, indicating a
continuous phase transition. Here we have used an exponent
$\alpha/\nu = 0.175$ corresponding to the 3D Ising universality. To
further characterize the phase transition, we turn to the
finite-size scaling of the order parameter.

\begin{figure} [t]
\includegraphics[width=1.0\columnwidth]{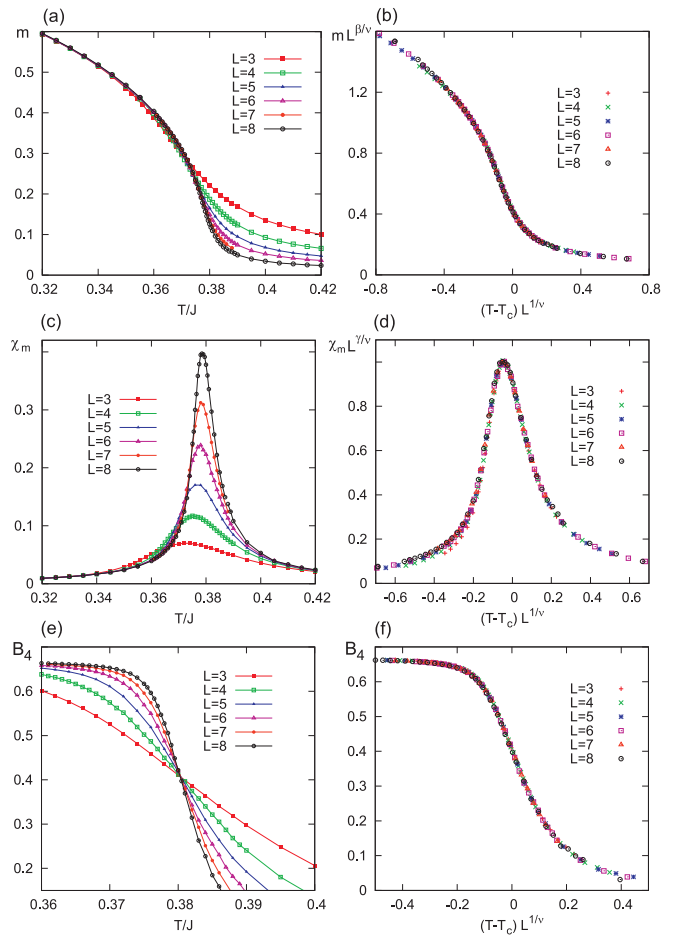}
\caption{\label{fig:d-dm2} The temperature dependence of (a) order
parameter $|m|$, (c) susceptibility $\chi_m$, and (e) Binder's
cumulant $B_{4m}$ obtained from Monte Carlo simulations for the
specific case of $D = 0.14J$. The right panels (b), (d), and (f)
show the corresponding finite-size scaling plots. The estimated
critical exponents are $\beta = 0.38$, $\gamma = 1.121$ and $\nu =
0.63$.}
\end{figure}

In the all-in all-out structure, spins point along the local
$\langle 111 \rangle$ directions with tetrahedra of opposite
orientations in the all-in and all-out states, respectively. We
introduce an Ising order parameter to characterize this doubly
degenerate state:
\begin{eqnarray}
    \label{eq:ising-op}
    m = \frac{1}{NS}\sum_i \mathbf S_i\cdot\hat\mathbf e_i,
\end{eqnarray}
where $N$ is the number of spins and $\hat\mathbf e_i$ points along
the local $[111]$ axes, i.e. from the center of type-I tetrahedron
to the corresponding corner.  The magnetic susceptibility and
Binder's cumulant are defined accordingly
\begin{eqnarray}
    \chi_m = \frac{N}{T} \left(\langle m^2\rangle - \langle
    |m|\rangle^2\right), \quad
    B_{4m} = 1 - \frac{\langle m^4 \rangle}{3\langle m^2 \rangle^{\,2}}.
\end{eqnarray}
The temperature dependence of these variables are shown in
Fig.~\ref{fig:d-dm2} for various linear size $L$. As can be seen
from Fig.~\ref{fig:d-dm2}(a), the order parameter $|m|$ starts to
grow below a critical temperature $T_c \approx 0.38J$ estimated from
the crossing of the Binder's cumulant shown in
Fig.~\ref{fig:d-dm2}(e).

The finite-size scaling plots shown in Figs.~\ref{fig:d-dm2}(b),
(d), and (f) indicate that the data points indeed collapse on a
universal curve. The critical exponents of the phase transition can
be estimated from the corresponding scaling relations. For example,
scaling of Binder's cumulant $B_{4m} =
\mathcal{B}((T-T_c)L^{1/\nu})$ yields the critical exponent $\nu =
0.63$. Combined with the ratio $\alpha/\nu$ obtained from the
specific heat, we find $\alpha = 0.11$. Similarly, from the scaling
of $|m|$ and $\chi_m$, we obtain critical exponents $\beta = 0.38$
and $\gamma = 1.121$.

The discrete $Z_2$ symmetry of order parameter $m$ implies a phase
transition in the 3D Ising universality class. We found that the
critical exponents $\alpha$ and $\nu$ indeed agree with the expected
values for Ising transitions in three dimensions.  The exponents
$\beta$ and $\gamma$, however, differ from the corresponding values
of the 3D Ising class ($\beta = 0.32$ and $\gamma = 1.24$).
\cite{chaikin} As discussed in Sec.~\ref{sec:intro}, this
discrepancy could be caused by the dipolar-like effective
interaction between spins in the Coulomb phase. Nonetheless, the
obtained exponents roughly satisfy the identity $\alpha + 2\beta +
\gamma \approx 2$.

In Appendix~\ref{app:dm}, we numerically study the magnetic ordering
of Hamiltonian~(\ref{eq:Hdm}) assuming $J = 0$. As discussed in the
previous section, the same all-in all-out magnetic order is selected
as the ground state. Finite-size scaling analysis with the order
parameter $m$ shows that the magnetic phase transition indeed
belongs to the 3D Ising universality class in the $J = 0$ limit.
This contrasting result demonstrates that the discrepancy observed
in exponents $\beta$ and $\gamma$ is attributed to the constraints
imposed by a large $J$ on the motion of spins in the Coulomb phase.

\begin{figure}
\includegraphics[width=1.\columnwidth]{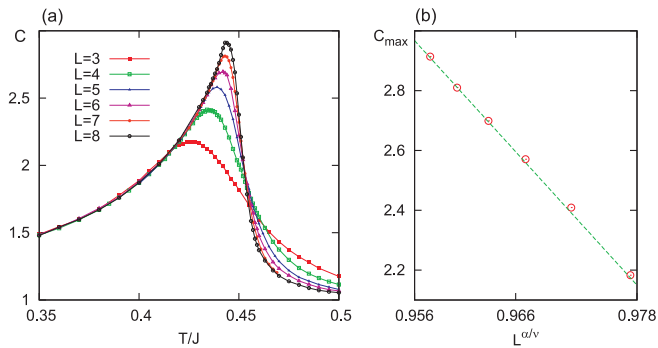}
\caption{\label{fig:id-dm1} Monte Carlo simulations for coupling
constants $D = -0.7J$. (a) Specific heat $C = \left(\langle
E^2\rangle - \langle E\rangle^2\right)/NT^2$ as a function of
temperature $T$. (b) The maximum of the specific heat $C_{\rm max}$
as a function of $L^{\alpha/\nu}$, where $\alpha/\nu = -0.0217$.}
\end{figure}

\section{XY-like phase transition: indirect DM interaction}
\label{sec:indirect-dm}

The indirect DM interaction only partially removes the magnetic
frustration. At the single tetrahedron level, the degenerate
ground-state manifold is parameterized by either $\phi$ or $\theta$
introduced in Eqs.~(\ref{eq:orthogonal}) and
(\ref{eq:non-coplanar}), respectively. Nonetheless, the system
undergoes a continuous phase transition at a temperature $T_c \sim
\mathcal{O}(|D|)$. A clear peak can be seen in the temperature
dependence of the specific heat as shown in
Fig.~\ref{fig:id-dm1}(a). To characterize the phase transition, we
need an order parameter which transforms as an irreducible
representation of the symmetry group and properly describes the
ground-state manifold.

\begin{figure} [t]
\includegraphics[width=0.9\columnwidth]{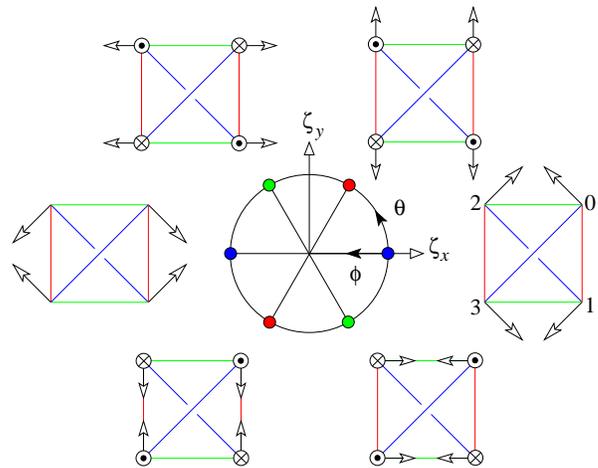}
\caption{\label{fig:circle} The domain of the order parameter
$\bm\zeta=(\zeta_x, \zeta_y)$ is bounded by a circle whose
circumference corresponds to noncoplanar ground states described by
Eq.~(\ref{eq:non-coplanar}). The coplanar ground states given in
Eq.~(\ref{eq:orthogonal}) lies on the three straight lines $\zeta_y
= \pm\sqrt{3}\zeta_x$ and $\zeta_y = 0$. Also shown are the six
orthogonal magnetic structures where spins point along $\langle
110\rangle$ directions; The $\odot$ and $\otimes$ symbols denote
spin component coming out of and into the plane, respectively. They
correspond to angles $\theta_{\zeta} = n\pi/3$ where $n = 0, \cdots,
5$.}
\end{figure}

To this end, we note that a general ground state of the indirect DM
model can be constructed using the three orthogonal structures in
which spins point along the $\langle 110\rangle$ directions. We thus
introduce unit vectors $\hat\mathbf m^{c}_{0,3} = (\mp 1, \pm 1,
0)/\sqrt{2}$ and $\hat\mathbf m^{c}_{1,2} = (\mp 1, \mp 1,
0)/\sqrt{2}$ for coplanar spins perpendicular to the $c$ axis.
Similar vectors $\hat\mathbf m^a_i$ and $\hat\mathbf m^b_i$ can be
defined for the other two coplanar states. For a given spin
configuration $\{\mathbf S_i\}$, the projection onto these
orthogonal structures is given by
\begin{eqnarray}
    p_{\mu} = \frac{1}{NS}\sum_i \mathbf S_i\cdot\hat\mathbf m^{\mu}_i,
    \quad\quad \mu = a, b, c.
\end{eqnarray}
An order parameter $q = \max_{\mu} p_{\mu}^2$ was introduced in
Ref.~\onlinecite{canals08} to characterize the proximity of spin
configurations to one of the six orthogonal structures shown in
Fig.~\ref{fig:circle}. Their Monte Carlo simulations demonstrated
that $q$ indeed approaches its maximum $q_{\rm max} = 1$ as $T \to
0$, indicating that the ground state consists of tetrahedra with
orthogonal spins.

As already mentioned in Sec.~\ref{sec:intro}, the ordered phase has
two distinct types of broken $Z_6$ symmetry. To characterize the
different symmetry-breaking patterns, we introduce a two-component
order parameter:
\begin{eqnarray}
    \zeta_x = \left(p_a + p_b - 2p_c\right)/\sqrt{6}, \quad
    \zeta_y = \left(p_a - p_b\right)/\sqrt{2},
\end{eqnarray}
which transforms as the doublet representation of the group $O_h$.
Note that the three projections are not independent as their
symmetric sum $p_a + p_b + p_c$ vanishes identically. The domain of
possible values of the vector $\bm\zeta = (\zeta_x, \zeta_y)$ is
bounded by a circle with radius $\sqrt{3/2}$
[Fig.~\ref{fig:circle}]. Its circumference is made of noncoplanar
ground states which are parameterized by $\theta$
[Eq.~(\ref{eq:non-coplanar})]:
\begin{eqnarray}
    \zeta_x =
    \frac{\sqrt{3}\cos\theta}{\sqrt{2+4\sin^2\theta}}, \quad
    \zeta_y =
    \frac{3\sin\theta}{\sqrt{2+4\sin^2\theta}}.
\end{eqnarray}
Noting that $\zeta_y/\zeta_x =\tan\theta_{\zeta} =
\sqrt{3}\tan\theta$, the six orthogonal structures shown in
Fig.~\ref{fig:circle} correspond to $\theta_{\zeta} = n\pi/3$ on the
circumference, where $n = 0,\cdots 5$. These six states are the
magnetic ground state at $T \to 0$.

Another set of degenerate ground states is given by the three
straight lines $\zeta_y = \pm \sqrt{3} \zeta_x$ and $\zeta_y = 0$.
They correspond to the coplanar spins described in
Eq.~(\ref{eq:orthogonal}); the order parameter of an orthogonal
structure parameterized by $\phi$ is given by
\begin{eqnarray}
    \label{eq:coplanar-z}
    \zeta_x = \sqrt{3/2}\cos\eta_{\mu}\cos\phi,\quad \zeta_y
    = \sqrt{3/2}\sin\eta_{\mu}\cos\phi,
\end{eqnarray}
where $\eta_{a/b} = \pm 2\pi/3$ and $\eta_c =0$, and $\phi$
describes the global rotation of the coplanar spins about the
corresponding cubic axes.

\begin{figure}
\includegraphics[width=1.0\columnwidth]{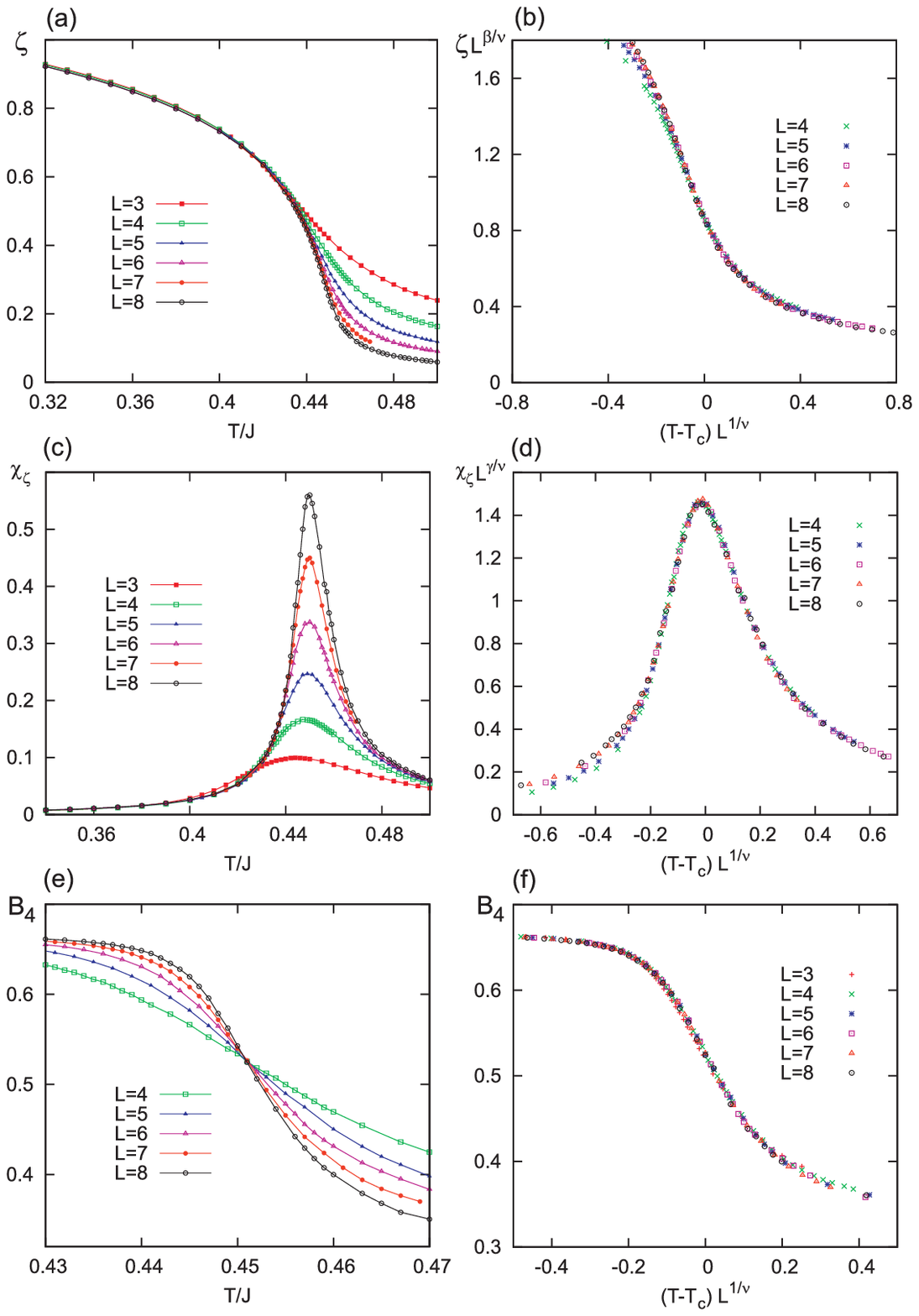}
\caption{\label{fig:id-dm2} The temperature dependence of (a) order
parameter $\zeta = |\bm\zeta|$, (c) susceptibility $\chi_{\zeta}$,
and (e) Binder's cumulant $B_{4\zeta}$ obtained from Monte Carlo
simulations for the specific case of $D = -0.7J$. The right panels
(b), (d), and (f) show the corresponding finite-size scaling plots.
The estimated critical exponents are $\beta = 0.42$, $\gamma =
1.181$, and $\nu = 0.672$.}
\end{figure}

Fig.~\ref{fig:id-dm2}(a) shows the order parameter  as a function of
temperature obtained from Monte~Carlo simulations. The amplitude
$\langle |\bm\zeta|\rangle$ of the order parameter increases
markedly below a temperature $T_c \approx 0.45J$. In
Figs.~\ref{fig:id-dm2}(c) and (e), we show the temperature
dependence of the susceptibility and Binder's cumulant defined as
\begin{eqnarray}
    \chi_{\zeta} = \frac{N}{T}\left(\langle |\bm\zeta|^2 \rangle -
    \langle |\bm\zeta|\rangle^2\right), \quad
    B_{4\zeta} = 1 - \frac{\langle |\bm\zeta|^4\rangle}{3\langle
    |\bm\zeta|^2\rangle^{\,2}}.
\end{eqnarray}
Again, a clear peak which diverges with the system size can be seen
in the susceptibility curves. The crossing of Binder's cumulants at
a temperature $T_c \approx 0.452 J$ indicates that the phase
transition is second-order.

The order parameter $\bm\zeta$ possesses a symmetry similar to that
of XY model with a $Z_6$ anisotropy term. It is known that this
anisotropy perturbation is dangerously irrelevant in 3D and the
ordering transition is in the 3D XY universality class.
\cite{jose,scholten,oshikawa,hove03} To investigate the nature of
the observed phase transition, we performed finite-size scaling on
the relevant quantities. The critical exponent $\nu$ which
characterizes the scaling of correlation length can be estimated
from the finite-size scaling plot of Binder's cumulant. As
Fig.~\ref{fig:id-dm2}(f) shows, using $\nu = 0.672$ of the XY
universality gives a rather good data collapsing. On the other hand,
it is known that the XY model has a small negative exponent $\alpha
= -0.0146$. The nonzero regular component of the specific heat makes
its finite-size scaling a rather difficult task.
Fig.~\ref{fig:id-dm1}(b) shows the maximum of the specific heat as a
function of scaled system size: $C_{\rm max} = C_0 + {\rm
const}\times L^{\alpha/\nu}$. Using $\alpha/\nu = -0.0217$ from the
XY universality class yields an agreeable result for large $L$.

The finite-size scaling plot of the order parameter
[Fig.~\ref{fig:id-dm2}(b)] shows that the data points indeed fall on
a universal curve $\zeta = L^{-\beta/\nu}\Phi((T-T_c)L^{1/\nu})$,
from which we estimate the critical exponent $\beta = 0.42$.
Similarly, scaling analysis of the susceptibility using relation
$\chi_{\zeta} = L^{-\gamma/\nu}\Upsilon((T-T_c) L^{1/\nu})$ yields
$\gamma = 1.181$. These two exponents differ from the expected
values of 3D XY universality class ($\beta = 0.3485$ and $\gamma =
1.3177$). \cite{xy} As already discussed in the case of direct DM
interaction, the discrepancy could be due to the constrained
fluctuations of spins in the Coulomb phase. The long-range dipolar
correlation of spins in this phase could modify the expected XY
critical behavior. Nonetheless, the deviations in $\beta$ and
$\gamma$ roughly compensate each other such that the critical
exponents still satisfy the Rushbrooke equality $\alpha + 2\beta +
\gamma \approx 2$. \cite{chaikin}

\section{Order-by-disorder and Crossover phenomenon}
\label{sec:crossover}

\subsection{Crossover between two types of $Z_6$ order}

Although the order parameter $\bm\zeta$ is indeed nonzero in the
ordered phase, the pattern of the broken symmetry is unclear from
the above Monte Carlo simulations. Expecting a state with broken
$Z_6$ symmetry, we define another order parameter which is sensitive
to the angular distribution of the vector $\bm\zeta$:
\cite{heilmann,lou07}
\begin{eqnarray}
    \zeta_6 = \langle |\bm\zeta|\cos(6\theta_{\zeta})\rangle,
\end{eqnarray}
where $\theta_{\zeta} = \arctan(\zeta_y/\zeta_x)$. The coplanar
ground states with $\theta_{\zeta} = n\pi/3$ is characterized by a
positive $\zeta_6$. In particular, a maximum $\zeta_{6,{\rm max}} =
\sqrt{3/2}$ indicates that the system is in one of the six
orthogonal structures shown in Fig.~\ref{fig:circle}. We refer to
this symmetry-breaking pattern as the type-I order. Another type of
broken $Z_6$ symmetry, referred to as the type-II order, corresponds
to order parameter $\bm\zeta$ clustered around the six angular
positions $\theta_{\zeta} = (n+1/2)\,\pi/3$. Tetrahedra in the
type-II order consist of noncoplanar spins shown in
Fig.~\ref{fig:tetra}(c). The $Z_6$ order parameter is negative
$\zeta_6 < 0$ for the type-II order.

\begin{figure}
\includegraphics[width=0.95\columnwidth]{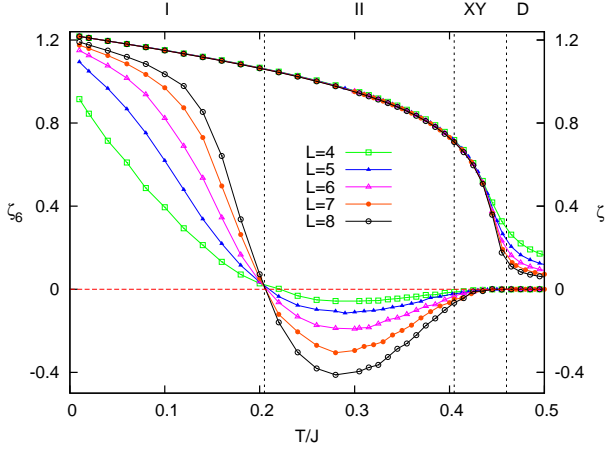}
\caption{\label{fig:m6} Temperature dependence of the XY order
parameter $\zeta = |\bm\zeta|$ and $Z_6$ order parameter $\zeta_6$
for system size $L = 4,5,6,7,8$. The instantaneous distribution of
the order parameter $\bm\zeta$ in the four distinct regimes D, XY, I
and IV are shown in Fig.~\ref{fig:hist}.}
\end{figure}

\begin{figure}[t]
\includegraphics[width=0.97\columnwidth]{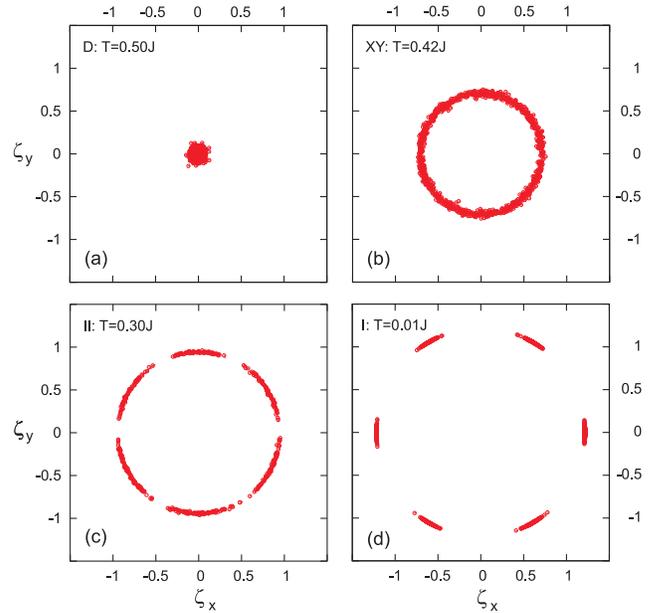}
\caption{\label{fig:hist} Distribution of the instantaneous order
parameter on the $(\zeta_x,\zeta_y)$ plane for (a) disordered phase,
(b) rotationally symmetric regime, (c) type-II and (d) type-I phases
with broken $Z_6$ symmetry.}
\end{figure}

Fig.~\ref{fig:m6} shows the temperature dependence of the two order
parameters $\zeta_6$ and $\zeta =\langle |\bm\zeta|\rangle$ obtained
from Monte~Carlo simulations. Depending on the angular distribution
of the doublet vector $\bm\zeta$ (shown in Fig.~\ref{fig:hist}), we
divide the ordered phase into three distinct regimes. In the XY
regime just below $T_c$, the distribution of the order parameter
exhibits an emergent rotational symmetry [Fig.~\ref{fig:hist}(b)].
This result is consistent with numerical studies on the closely
related 3D six-clock model which confirmed the irrelevance of the
anisotropy at $T_c$.

Surprisingly, instead of directly settling in the type-I order which
is shown to be the ground state as $T\to 0$ in
Ref.~\onlinecite{canals08}, the system first enters a regime with
type-II order as indicated by a negative $\zeta_6$. Upon further
lowering the temperature, the system gradually switches to the
type-I order. A crossover temperature can be estimated from the
point where the order parameter $\zeta_6$ changes sign. At $T \to 0$
the parameter $\zeta_6$ approaches its maximum indicating a ground
state with orthogonal spins.

As discussed in Sec.~\ref{sec:indirect-dm}, the continuous
degeneracy $\theta$ of the noncoplanar ground state corresponds to a
rotational symmetry of the doublet vector $\bm\zeta$. As a result,
the observed broken $Z_6$ symmetry, either the type-I or type-II
order, must be of entropic origin. Naively, one would expect quantum
or thermal fluctuations select the six orthogonal structures shown
in Fig.~\ref{fig:circle} because they possess two zero modes
corresponding to directions $\theta$ and $\phi$ in the ground-state
manifold. By considering zero-point energy $\sum_m \varepsilon_m/2$
of magnons, Canals {\em et al.} showed that quantum fluctuations
indeed favor the six orthogonal structures. However, at finite
temperatures one must consider the magnon contribution to the system
free energy. As we demonstrate below, the high-energy magnons
actually play an important role in the high-temperature regime of
the ordered phase.

\subsection{Order from disorder: Holestein-Primakoff expansion}

Here we compute the free energy due to the harmonic fluctuations
around a given noncoplanar magnetic order. The free energy is a
function of $\theta$ which parameterizes the $\mathbf q=0$
noncoplanar spins given in Eq.~(\ref{eq:non-coplanar}). We first
introduce a local reference frame defined by three orthonormal
vectors: $\hat{\bf e}^x_i$, $\hat{\bf e}^y_i$, and $\hat{\bf n}_i$,
which are functions of $\theta$. A small deviation from the
noncoplanar structure can be expressed using the Holstein-Primakoff
expansion:
\begin{equation}
    {\bf S}_i = S \Bigl(1-\frac{\bm{\sigma}_i^2}{2 S^2}\Bigr)\,\hat{\bf n}_i
    +\sum_{a=x,y} \sigma^a_i\, \hat{\bf e}^a_i
    + \mathcal{O}(\sigma^3).
\end{equation}
Here $\bm\sigma_i = (\sigma^x_i, \sigma^y_i)$ whose components
denote fluctuations along the two orthogonal local axes.
Substituting the above expression into Eq.~(\ref{eq:Hdm}), we obtain
a magnon Hamiltonian
\begin{eqnarray}
    \mathcal{H} = \bigl(J-\sqrt{2}D\bigr)\sum_i |\bm\sigma_i|^2 +
    \sum_{i\neq j}\sum_{a,b=x,y}H^{ab}_{ij}\sigma^a_i\sigma^b_j,
\end{eqnarray}
where
\begin{eqnarray}
    H^{ab}_{ij} = \frac{1}{2}\left(J\hat\mathbf
    e^a_i\cdot\hat\mathbf e^b_j + \mathbf D_{ij}\cdot\hat\mathbf
    e^a_i\times\hat\mathbf e^b_j\right).
\end{eqnarray}
Since the noncoplanar structure is translationally invariant
$(\mathbf q = 0)$, the above Hamiltonian is diagonalized by Fourier
transformation. In the vicinity of the coplanar state $\theta_\zeta
= \theta = 0$, the lowest energy band has a gapless dispersion
\begin{eqnarray}
    \varepsilon_{1,\mathbf k} \approx
    \frac{1}{12}(2J+5\sqrt{2}D) \left| \mathbf k_{\perp}\right|^2 +
    \frac{1}{6}(4J+\sqrt{2}D) k_z^2,
\end{eqnarray}
where $|\mathbf k_{\perp}|^2 = k_x^2 + k_y^2$. The zero mode at
$\mathbf k = 0$ corresponds to the direction $\theta$ in the
ground-state manifold [see Fig.~\ref{fig:circle}]. The dispersion of
the second lowest band is
\begin{eqnarray}
    \varepsilon_{2,\mathbf k} \approx \Delta + \frac{1}{4}
    (2J+\sqrt{2}D) \left|\mathbf k_{\perp}\right|^2+
    \frac{D}{\sqrt{2}} k_z^2,
\end{eqnarray}
where the energy gap is due to a finite DM coefficient and is
proportional to the deviation $\theta$ from the coplanar state:
\begin{eqnarray}
    \Delta \approx 4\sqrt{2}D\,\theta^2.
\end{eqnarray}
The vanishing of the gap at $\theta = 0$ corresponds to the zero
mode described by the $\phi$ direction in Fig.~\ref{fig:circle},
i.e. uniform rotation of spins about the cubic axes. At low
temperatures, spin fluctuations are dominated by the low energy
magnons. Due to the presence of two zero modes at $\theta = \phi =
0$, the six orthogonal structures shown in Fig.~\ref{fig:circle} are
selected as the ground state as $T \to 0$, consistent with the
Monte~Carlo simulations.

\begin{figure}
\includegraphics[width=0.85\columnwidth]{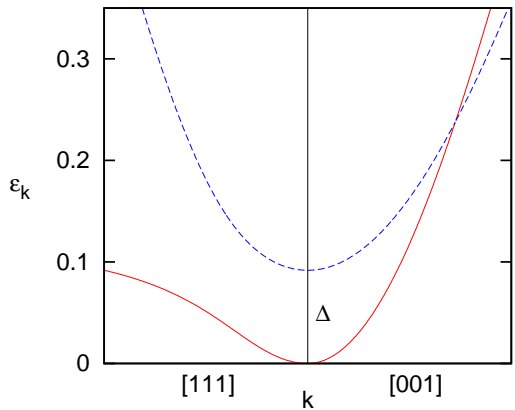}
\caption{\label{fig:ek} Dispersion of the lowest two energy bands
near the $\Gamma$ point. The gapless band contains a zero mode at
$\mathbf k = 0$ corresponding to the direction $\theta$ in the
ground state manifold [c.f. Fig.~\ref{fig:circle}]. The second band
is gapped with $\Delta \sim D\,\theta^2$.}
\end{figure}

\begin{figure}
\includegraphics[width=0.75\columnwidth]{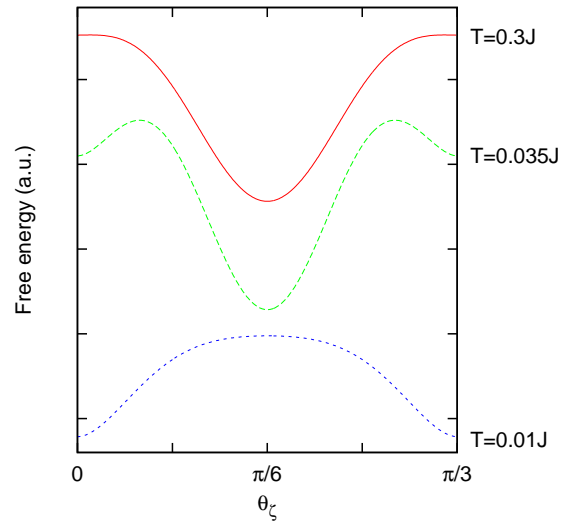}
\caption{\label{fig:f-t} Magnon free energy Eq.~(\ref{eq:f-e}) as a
function of angle $\theta_{\zeta} = \arctan(\zeta_y/\zeta_x)$. Note
that this angle is related to the parameter $\theta$ characterizing
a general noncoplanar structure [Eq.~(\ref{eq:non-coplanar})]
through $\tan\theta_{\zeta} = \sqrt{3}\tan\theta$. The coupling
constant used in the calculation is $D = -0.2J$. Note that
$\theta_{\zeta} = \pi/6$ corresponds to noncoplanar structure shown
in Fig.~\ref{fig:tetra}(c), whereas $\theta_{\zeta} = 0$ and $\pi/3$
correspond to orthogonal spins shown in Figs.~\ref{fig:circle} and
\ref{fig:tetra}(b).}
\end{figure}

At higher temperatures, one needs to take into account the
contributions from the high energy bands. The magnon contribution to
the free energy is given by
\begin{eqnarray}
    \label{eq:f-e}
    F_{\rm magnon}(\theta_{\zeta}) = \sum_{n, \mathbf k} T
    \ln\sinh\left[\,\varepsilon_{n,\,\mathbf k}(\theta_{\zeta})/T\right].
\end{eqnarray}
The angle $\theta_{\zeta} = \arctan(\zeta_y/\zeta_x)$ is related to
the parameter $\theta$ through $\tan\theta_{\zeta} =
\sqrt{3}\tan\theta$. We numerically compute the free energy by
summing over states within the Brillouin zone. Fig.~\ref{fig:f-t}
shows the free energy as a function of $\theta_{\zeta}$. At higher
temperatures, e.g. $T = 0.3J$, the minimum of $F_{\rm magnon}$
occurs at $\theta_{\zeta} = (n+1/2)\pi/3$ corresponding to
noncoplanar spins shown in Fig.~\ref{fig:tetra}(c). As the
temperature decreases, local minimum starts to develop at angles
$\theta_{\zeta} = n\pi/3$ which correspond to the six orthogonal
structures shown in Fig.~\ref{fig:circle} or
Fig.~\ref{fig:tetra}(b). As temperature is further lowered, the
orthogonal magnetic order takes over and becomes the absolute
minimum of the free energy. This calculation clearly shows that
thermal fluctuations favor two distinct spin structures at different
temperatures of the ordered phase, as indeed observed in Monte~Carlo
simulations.

\section{conclusion}
\label{sec:conclusion}

To summarize, we have studied the critical behavior of a classical
pyrochlore antiferromagnet with Dzyaloshinskii-Moriya interactions.
Symmetry considerations allow for two distinct forms of
antisymmetric coupling on the pyrochlore lattice which are termed
direct and indirect DM interactions. \cite{elhajal05,canals08} The
macroscopic degeneracy of the nearest-neighbor Heisenberg spins is
lifted for both types of DM interaction. Through extensive Monte
Carlo simulations, we have demonstrated the continuous nature of
magnetic ordering in both cases. The critical exponents of the phase
transitions were obtained from finite-size scaling analysis.
Finally, we have shown that the ordered phase in the indirect DM
model exhibits two distinct types of broken $Z_6$ symmetry.

The direct DM interaction completely relieves the geometrical
frustration and selects a doubly degenerate all-in all-out magnetic
order. We have observed an Ising-like phase transition with critical
exponents $\alpha = 0.11$, $\beta = 0.38$, $\gamma = 1.121$, and
$\nu = 0.63$. The deviation from the expected 3D Ising universality
is attributed to the correlated spin fluctuations in the Coulomb
phase. This is because minimization of the nearest-neighbor exchange
requires the total spin on every tetrahedra be zero, which in turn
results in a dipolar-like spin correlation at large distances. As is
well known, the long-range dipolar spin interaction could modify the
nature of the magnetic phase transition.

We have also simulated a pyrochlore magnet where the only coupling
between spins is the nearest-neighbor DM interaction. Noting that
this model selects exactly the same all-in all-out ground state, we
have shown that the corresponding magnetic transition indeed belongs
to the 3D Ising universality class when spin fluctuations are not
subject to local constraints imposed by the nearest-neighbor
exchange.

To characterize the symmetry-breaking phase of the indirect DM
model, we have introduced a doublet order parameter $\bm\zeta$. The
domain of this doublet vector is bounded by a circle whose
circumference is composed of degenerate noncoplanar ground states.
On the other hand, coplanar ground states with spins perpendicular
to one of the cubic axes lie on the lines $\zeta_y = \pm
\sqrt{3}\zeta_x$ and $\zeta_y = 0$. Finite-size scaling analysis
using the doublet order parameter yields a set of critical
exponents: $\alpha = -0.0146$, $\beta = 0.42$, $\gamma = 1.181$, and
$\nu = 0.672$. Similar to the case of direct DM interaction, we have
observed noticeable deviations in $\beta$ and $\gamma$ from the
expected 3D XY universality class. Again, the dipolar spin
correlation in the Coulomb phase can cause the observed discrepancy.

Our simulations have also uncovered an unusual order from disorder
phenomenon in the case of indirect DM interaction. Thermal
fluctuations below the transition temperature lift the continuous
degeneracy and select a $\mathbf q=0$ magnetic order with six-fold
degeneracy. Interestingly, the magnet first develops a noncoplanar
magnetic order before settling in the coplanar ground state. The
crossover between these two types of $Z_6$ order corresponds to a
rotation of the doublet vector $\bm\zeta$ from $\theta_\zeta =
(n+1/2)\pi/3$ to $\theta_\zeta = n\pi/3$, where $n$ is an integer.
By computing the magnon contribution to the free energy, we have
explicitly demonstrated that the observed crossover indeed
originates from the entropic selection.

\begin{acknowledgments}
I thank E. Choi, N. Perkins, C. Fennie, and O. Tchernyshyov for
collaboration on related works.
\end{acknowledgments}

\appendix

\section{Continuum approximation}
\label{app:continuum}

Here we outline a continuum description of the pyrochlore
antiferromagnet with DM interaction. Following
Refs.~\onlinecite{isakov04} and \onlinecite{henley05}, we define a
magnetic or polarization field at the center of a tetrahedron as
\begin{eqnarray}
    \mathbf B^a = \sum_{i\in\boxtimes}
    S^a_i\,\hat\mathbf e_i,
\end{eqnarray}
where $\hat\mathbf e_i$ denotes the local $\langle 111 \rangle$
direction at site $i$. Comparing with the definition of the three
staggered magnetizations, we observe
\begin{eqnarray}
    \mathbf B^a =4S (L^a_1, L^a_2, L^a_3).
\end{eqnarray}
This relation was first noted in Ref.~\onlinecite{colon}. In the
coarse-grained approximation, the constraint $\mathbf M = 0$
translates to $\nabla\cdot\mathbf B^a(\mathbf r) = 0$ for each
component $a$. Expressed in terms of the staggered fields, we have
\begin{eqnarray}
    \label{eq:div-free}
    \partial_x L^a_1 + \partial_y L^a_2 + \partial_z L^a_3 = 0.
\end{eqnarray}
Noting that the states with small values of $\mathbf B^a$ are
entropically favored, the probability distribution of the flux field
has a Gaussian distribution, i.e. $\rho \propto
e^{-\mathcal{H}_{2a}}$ with
\begin{eqnarray}
    \mathcal{H}_{2a} = \frac{\kappa}{2}\int d^3\mathbf r
    \sum_i \left|\mathbf L_i(\mathbf r)\right|^2.
\end{eqnarray}
The stiffness $\kappa$ is the single parameter of the theory which
controls the amplitude of the correlations. The correlators of the
staggered magnetizations at large distances are
\begin{eqnarray}
    \langle L^a_i(\mathbf r)\,L^b_j(0)\rangle \propto
    \frac{\delta_{ab}}{\kappa}\,\frac{r^2 \delta_{ij} - 3 r_i r_j}{r^5}.
\end{eqnarray}
While $\mathcal{H}_{2a}$ is entirely of entropic origin, an
additonal energy term comes from the DM interaction
\begin{eqnarray}
    \mathcal{H}_{2b} &=& -\mathcal{D} \int d^3\mathbf r \, \bigl[\,\hat\mathbf
    a\cdot\mathbf L_2(\mathbf r)\times\mathbf L_3(\mathbf r) \\
    & & +\hat\mathbf b\cdot\mathbf L_3(\mathbf r)\times\mathbf L_1(\mathbf r)
    +\hat\mathbf c\cdot\mathbf L_1(\mathbf r)\times\mathbf
    L_2(\mathbf r)\bigr], \nonumber
\end{eqnarray}
where the coupling constant $\mathcal{D} \propto DS^2$. The
competition between the two terms leads to a phase transition into
an ordered phase with nonzero $\langle \mathbf L_i(\mathbf
r)\rangle$. Details of the staggered fields are discussed in
Sec.~\ref{sec:dm} for the two different forms of DM interaction.

To ensure stability of the ordered phase and give penalties to
short-wavelength fluctuations, we add gradient and quartic terms
\begin{eqnarray}
    \mathcal{H}_{2c} = \int d^3\mathbf r \sum_i \Bigl[A_1
    \left|\nabla\cdot\mathbf L_i(\mathbf r)\right|^2 +
    A_2 \left|\nabla\times\mathbf L_i(\mathbf
    r)\right|^2\Bigr],\quad
\end{eqnarray}
\begin{eqnarray}
    \mathcal{H}_4= \int d^3\mathbf r \Bigl[ u\sum_i \left|\mathbf L_i(\mathbf
    r)\right|^4 + v\sum_{i\neq j} \left|\mathbf L_i(\mathbf
    r)\right|^2 \left|\mathbf L_j(\mathbf
    r)\right|^2\Bigr].\quad
\end{eqnarray}
The sum $\mathcal{H}_2 + \mathcal{H}_4$ thus constitute a continuum
framework for describing the phase transition induced by DM
interactions. Although the resultant expression resembles a
conventional Landau-Ginzburg-Wilson functional expressed in terms of
the antiferromagnetic order parameters, it is important to note that
these order parameter fields are not indepenent. Instead, they are
subject to the constraint~(\ref{eq:div-free}). This condition can be
satisfied by introducing a vector field $\mathbf A^a = (A^a_1,
A^a_2, A^a_3)$ for each spin component $a$ such that $\mathbf B^a =
\nabla\times\mathbf A^a$. Regrouping the vector potentials
$\tilde\mathbf A_i = (A^x_i, A^y_i, A^z_i)$, the coarse-grained
staggered fields is given by
\begin{eqnarray}
    \mathbf L_i(\mathbf r) =
    \epsilon_{ijk}\,\partial_j \tilde\mathbf A_k(\mathbf r)
\end{eqnarray}
A detailed analysis of the resulting energy functional in terms of
$\tilde\mathbf A_k(\mathbf r)$ will be presented in future
publications.

\section{Ising transition in direct DM model with $\bm{J = 0}$}
\label{app:dm}

\begin{figure}[b]
\includegraphics[width=1.\columnwidth]{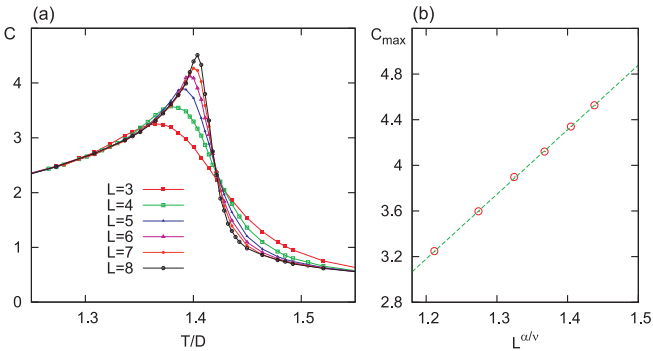}
\caption{\label{fig:d-dm1-j0} (a) Specific heat $C$ as a function of
temperature $T$ for various system sizes. (b) The maximum of the
specific heat $C_{\rm max}$ as a function of $L^{\alpha/\nu}$, where
$\alpha/\nu = 0.175$. The number of spins $N = 16 L^3$ for a system
with linear size $L$.}
\end{figure}

\begin{figure} [t]
\includegraphics[width=1.0\columnwidth]{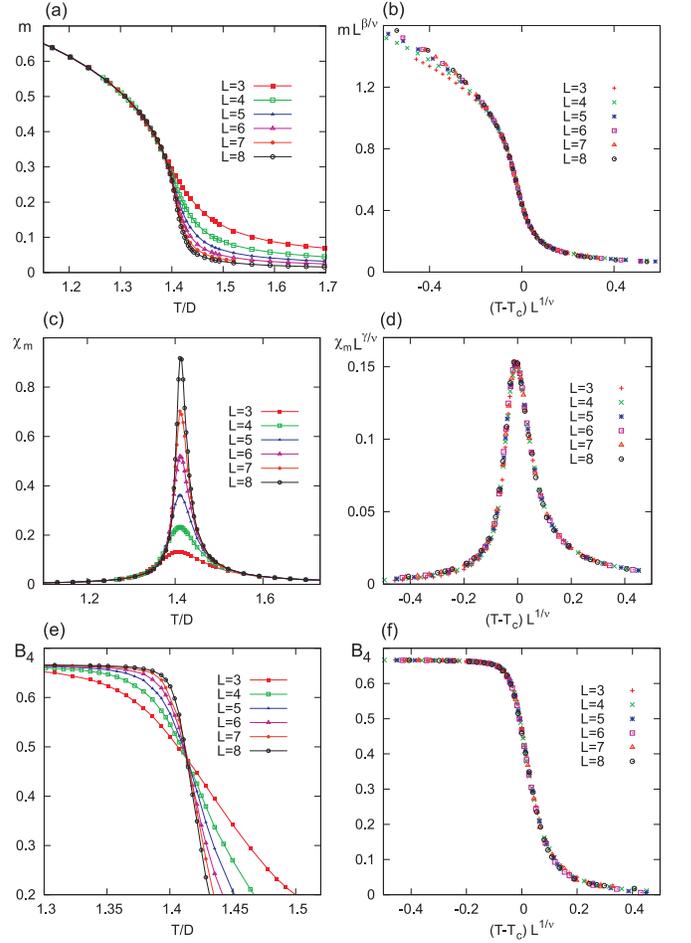}
\caption{\label{fig:d-dm2-j0} The temperature dependence of (a)
order parameter $|m|$, (c) susceptibility $\chi_m$, and (e) Binder's
cumulant $B_{4m}$ obtained from Monte Carlo simulations. The right
panels (b), (d), and (f) show the corresponding finite-size scaling
plots with 3D Ising critical exponents: $\beta = 0.32$, $\gamma =
1.24$ and $\nu = 0.63$.}
\end{figure}

In this appendix, we examine the critical behavior of classical
Heisenberg spins with direct DM interaction alone on the pyrochlore
lattice:
\begin{eqnarray}
    \mathcal{H} &=& \sum_{\langle ij \rangle} \mathbf D_{ij}\cdot
    \left(\mathbf S_i\times\mathbf S_j\right)
\end{eqnarray}
This model corresponds to the special case $J = 0$ of the
Hamiltonian~(\ref{eq:Hdm}) studied in the main text. Remarkably, the
anisotropic direct DM interaction alone selects the same all-in
all-out magnetic ground state [Fig.~\ref{fig:tetra}(a)] as in the
$J\to\infty$ limit. However, since spin fluctuations are not subject
to constraints $\sum_{i\in\boxtimes}\delta\mathbf S_i = 0$ imposed
by a large $J$, this model provides us an opportunity to examine how
the universality class of the magnetic transition is modified by the
long-range dipolar correlation of spins in the Coulomb phase.

Fig.~\ref{fig:d-dm1-j0}(a) shows the specific heat as a function of
temperature. A clear peak which diverges with increasing system
sizes indicates a continuous phase transition at $T_c \sim
\mathcal{O}(D)$.  Due to the existence of nonzero regular component
in the specific heat, finite-size scaling of $C$ is rather
difficult. Fig.~\ref{fig:d-dm1-j0}(b) shows the maximum of
specific-heat as a function of $L^{\alpha/\nu}$, where we have used
the ratio $\alpha/\nu = 0.175$ from 3D Ising universality class. The
data collapsing on a straight line indicates a scaling relation
$C_{\rm max} = C_0 + {\rm const}\times L^{\alpha/\nu}$. We use the
same order parameter $m$ defined in Eq.~(\ref{eq:ising-op}) to
characterize the doubly degenerate all-in all-out magnetic order.
The temperature dependence of $m$ and the corresponding
susceptibility and Binder's cumulant are shown in
Fig.~\ref{fig:d-dm2-j0}. Using critical exponents of 3D Ising
universality: $\beta = 0.32$, $\gamma = 1.24$, and $\nu = 0.63$, we
obtained excellent data collapsing as shown in the corresponding
finite-size scaling plots [Fig.~\ref{fig:d-dm2-j0}]. The analysis
shows that without the constraints imposed by nearest-neighbor
exchange $J$, the phase transition into the doubly degenerate all-in
all-out state indeed belongs to 3D Ising class.


\begin{thebibliography}{99}

\bibitem{moessner06} R. Moessner and A. P. Ramirez, Physics Today
{\bf 59} (2), 24 (2006).

\bibitem{moessner98} R. Moessner and J. T. Chalker,
Phys. Rev. Lett. {\bf 80}, 2929 (1998); Phys. Rev. B {\bf 58}, 12049
(1998).

\bibitem{hermele04} M. Hermele, M. P. A. Fisher, and L. Balents,
Phys. Rev. B {\bf 69}, 064404 (2004).

\bibitem{isakov04} S. V. Isakov, K. Gregor, R. Moessner, and
S. L. Sondhi, Phys. Rev. Lett. {\bf 93}, 167204 (2004).

\bibitem{henley05} C. L. Henley, Phys. Rev. B {\bf 71}, 014424 (2005).

\bibitem{lee00} S.-H. Lee, C. Broholm, T. H. Kim, W. Ratcliff II,
and S.-W. Cheong, Phys. Rev. Lett. {\bf 84}, 3718 (2000).

\bibitem{chung05} J.-H. Chung, M. Matsuda, S.-H. Lee, K. Kakurai, H.
Ueda, T. J. Sato, H. Takagi, K.-P. Hong, and S. Park, Phys. Rev.
Lett. {\bf 95}, 247204 (2005).


\bibitem{melko01} R. G. Melko, B. C. den Hertog, M. J. P. Gingras,
Phys. Rev. Lett. {\bf 87}, 067203 (2001).

\bibitem{cepas05} O. Cepas, A. P. Young, B. S. Shastry, Phys. Rev. B {\bf 72}, 184408
(2005).

\bibitem{pickles08} T. S. Pickles, T. E. Saunders, J. T. Chalker,
Europhys. Lett. {\bf 84} 36002 (2008).

\bibitem{saunders08} T. E. Saunders, J. T. Chalker, Physical Review B {\bf 77}, 214438
(2008).

\bibitem{tsuneishi07} D. Tsuneishi, M. Ioki, and H. Kawamura,
J. Phys. Condens. Matter, {\bf 19}, 145273 (2007).

\bibitem{chern08} G.-W. Chern, R. Moessner, and O. Tchernyshyov, Phys. Rev. B {\bf 78}, 144418
(2008).

\bibitem{brezin74} E. Br\'ezin, J. C. Le Guillou, and J.
Zinn-Justin, Phys. Rev. B {\bf 10}, 892 (1974).

\bibitem{bak76} P. Bak, S. Krinsky, and D. Mukamel, Phys. Rev. Lett.
{\bf 36}, 52 (1976).

\bibitem{larkin69} A. I. Larkin and D. E. Khmel'nitskii, Sov. Phys.
JETP {\bf 29}, 1123 (1969).

\bibitem{aharony73} A. Aharony, Phys. Rev. B {\bf 8}, 3363 (1973).

\bibitem{reimers91} J. N. Reimers, A. J. Berlinsky, and A.-C. Shi,
Phys. Rev. B {\bf 43}, 865 (1991).


\bibitem{reimers92} J. N. Reimers, J. E. Greedan, and M.
Bjorgvinsson, Phys. Rev. B {\bf 45}, 7295 (1992).

\bibitem{dz}
I. E. Dzyaloshinskii, Sov. Phys. JETP {\bf 19}, 960 (1964).

\bibitem{moriya} T. Moriya, Phys. Rev. {\bf 120,} 91 (1960).


\bibitem{chern06} G.-W. Chern, C. Fennie, and O. Tchernyshyov,
Phys. Rev. B {\bf 74}, 060405(R) (2006).

\bibitem{ot} O. Tchernyshyov and G.-W. Chern, in {\em Highly Frustrated Magnetism},
edited by C. Lacroix, P. Mendels, and F. Mila (Springer, 2010).

\bibitem{onose10} Y. Onose, T. Ideue, H. Katsura, Y. Shiomi, N.
Nagaosa, Y. Tokura, Science {\bf 329}, 297 (2010).


\bibitem{elhajal05} M. Elhajal, B. Canals, R. Sunyer, and C. Lacroix,
Phys. Rev. B {\bf 71}, 094420 (2005).

\bibitem{canals08} B. Canals, M. Elhajal, and C. Lacroix, Phys. Rev.
B {\bf 78}, 214431 (2008).

\bibitem{colon} P. H. Conlon, J. T. Chalker, Phys. Rev. B {\bf 81}, 224413
(2010).


\bibitem{chaikin} P. M. Chaikin and T. C. Lubensky, {\em Principles of
Condensed Matter Physics} (Cambridge University Press, Cambridge,
2000).

\bibitem{jose} J. V. Jose, L. P. Kadanoff, S. Kirkpatrick, and D. R.
Nelson, Phys. Rev. B {\bf 16}, 1217 (1977).

\bibitem{oshikawa} M. Oshikawa, Phys. Rev. B {\bf 61}, 3430 (2000).

\bibitem{scholten} P. D. Scholten and L. J. Irakliotis, Phys. Rev. B
{\bf 48}, 1291 (1993).

\bibitem{hove03} J. Hove and A. Sudbo, Phys. Rev. E {\bf 68}, 046107
(2003).

\bibitem{xy} M. Campostrini, M. Hasenbusch, A. Pelissetto, P. Rossi,
and E. Vicari, Phys. Rev. B {\bf 63}, 214503 (2001).

\bibitem{heilmann} R. K. Heilmann, J.-S. Wang, R. H. Swendsen, Phys.
Rev. B {\bf 53}, 2210 (1996).

\bibitem{lou07} J. Lou, A. W. Sandvik, L. Balents, Phys. Rev. Lett.
{\bf 99}, 207203 (2007).



\end{thebibliography}
\end{document}